\documentclass[aps,pre,onecolumn,superscriptaddress,nofootinbib]{revtex4-2}

\usepackage{amsmath,amssymb,amsfonts}
\usepackage{mathtools}
\usepackage{bm}
\usepackage[colorlinks=true,linkcolor=blue,citecolor=red]{hyperref}%
\usepackage{color}
\usepackage{tikz-cd}
\usepackage{comment}

\newcommand{\cM}{\mathcal{M}}
\newcommand{\cK}{\mathcal{K}}
\newcommand{\cR}{\mathcal{R}}
\newcommand{\cH}{\mathcal{H}}
\newcommand{\cL}{\mathcal{L}}

\def\x{\bm{x}}
\def\y{\bm{y}}
\def\q{q}
\def\ve{\varepsilon}
\def\ctanh{\mathrm{ctanh}}
\def\pa{\partial\Omega}

\begin{document}

\title{Encounter Propagator for Multiple Targets: \\ A Dirichlet-to-Neumann Spectral Formalism}

\author{Denis S. Grebenkov}
\affiliation{Laboratoire de Physique de la Mati\`ere Condens\'ee, CNRS -- \'Ecole Polytechnique, 
Institut Polytechnique de Paris, 91120 Palaiseau, France}

\date{\today}

\begin{abstract}
We develop an encounter-based formulation of restricted diffusion in a
bounded domain with multiple targets, resolving separately the
boundary local time accumulated on each target.  Unlike the
single-target problem, the target projection operators do not
generally commute with the governing Dirichlet-to-Neumann (DtN)
operator, so that its eigenbasis does not diagonalize the local-time
dependence.  We overcome this difficulty by decomposing the DtN
operator into target-restricted blocks.  Multi-dimensional Laplace
inversion then yields a convergent switching expansion, whose
successive terms describe alternating diffusive transfers between the
targets.  We also derive an equivalent operator-valued renewal
equation.  For two targets, diagonalizing the target-restricted blocks
provides an explicit spectral representation in terms of two DtN block
spectra and an inter-target coupling matrix.  When the coupling
preserves spectral modes, the switching series can be resummed exactly
in terms of modified Bessel functions.  In general, as the
off-diagonal DtN block is smoothing for separated targets, one can
resort to low-rank finite-dimensional approximations.  We examine an
effective two-mode reduction for small, well-separated targets and
identify a regime in which repeated inter-target transfers are
progressively suppressed.  Probabilistic interpretations and
implications for diffusion-controlled reactions are discussed.
\end{abstract}

\maketitle

\section{Introduction}

Diffusion-controlled reactions play an important role in various
physical, chemical and biological systems
\cite{North66,Wilemski73,Calef83,Berg85,Rice85,Grebenkov23f}, with
examples including nuclear magnetic relaxation \cite{Brownstein79},
ligand-protein associations \cite{Zwanzig90}, geminate recombination
of radicals and ions \cite{Sano79}, oxygen capture in the lungs
\cite{Weibel,Sapoval02}, to name but a few.  In a typical situation,
particles diffuse in the bulk toward partially reactive sites at the
boundary where they can react, relax an excited state, to absorbed or
destroyed.  The conventional theory of diffusion-controlled reactions
relies on the diffusion equation with mixed Robin-Neumann boundary
conditions.  In this way, the diffusive dynamics, governed by the
associated Laplace operator, is intrinsically coupled to the imposed
surface reactions.
In contrast, the encounter-based approach to diffusion-controlled
reactions aims at separating the diffusive exploration of a confining
domain from the reaction mechanism at its boundary
\cite{Grebenkov20}.  This separation is achieved by characterizing
surface encounters through boundary local time, while the reaction
mechanism is subsequently imposed as a stopping condition on that
local time.  The encounter-based approach found numerous applications
for describing the statistics of encounters between diffusing
particles \cite{Grebenkov21,Grebenkov22}, diffusion-mediated surface
reactions with stochastic resetting \cite{Bressloff22d,Benkhadaj22} or
with non-Markovian binding/unbinding kinetics \cite{Grebenkov23a},
probabilistic models of diffusion through semipermeable barriers
\cite{Bressloff22a,Bressloff22c,Bressloff23a,Bressloff23b}, escape
times from a bounded domain with reactive targets \cite{Grebenkov23b},
subdiffusive dynamics \cite{Bressloff23c,Bressloff23d}, the statistics
of adsorption and permeation events \cite{Grebenkov24a}.  The
inclusion of the boundary local time as a natural proxy for encounters
is the key distinction of this approach from conventional theories of
diffusion-controlled reactions and related first-passage problems
\cite{Redner,Schuss,Metzler,Lindenberg,Grebenkov,Bressloff13,Benichou14}.

The original formulation of the encounter-based formalism treated the
whole boundary $\pa$ of a confining domain $\Omega \subset {\mathbb
R}^d$ as homogeneously reactive.  When only a subset $\Gamma$ of the
boundary has a constant reactivity $q > 0$ (whereas the remaining
boundary $\Gamma_0 = \pa\backslash \Gamma$ is reflecting), the central
stochastic variable is the boundary local time $\ell_t$ accumulated on
that single reactive subset $\Gamma$ (see standard textbooks
\cite{Levy,Ito,Freidlin,Borodin} for its definition and properties).  A
probabilistic description of such diffusion-controlled processes
relies on the encounter propagator $P(\x,\ell,t|\x_0)$ -- the joint
probability density of the particle position $\bm{X}_t$ and its
boundary local time $\ell_t$ at time $t$, given that the starting
point was at $\x_0$.  Qualitatively, $P(\x,\ell,t|\x_0)$ describes the
propagation of the particle from $\x_0$ to $\x$ in time $t$,
accounting for the statistics of encounters with the subset $\Gamma$
via $\ell$.  The Laplace transform with respect to $\ell$,
\begin{equation}
G_q(\x,t|\x_0) = \int_0^\infty d\ell \, e^{-q\ell} \, P(\x,\ell,t|\x_0), \qquad q > 0,
\end{equation}
relates the encounter propagator to the heat kernel $G_q(\x,t|\x_0)$,
with a Robin boundary condition of parameter $q > 0$ on $\Gamma$ and
Neumann condition on $\Gamma_0$.  In this conventional formulation,
the heat kernel describes the diffusive propagation from $\x_0$ to
$\x$ in time $t$, with possible reaction on the partially reactive
target $\Gamma$.  An additional Laplace transform with respect to
time,
\begin{equation}
\widetilde G_q(\x,p|\x_0) = \int_0^\infty dt\, e^{-pt} \, G_q(\x,t|\x_0) , \qquad p > 0,
\end{equation}
transforms the heat kernel to the Robin-Neumann Green's function of
the modified Helmholtz equation: $(p - D\Delta) \widetilde
G_q(\x,p|\x_0) = \delta(\x-\x_0)$, where $D > 0$ is a constant
diffusion coefficient and $\delta(\x)$ is the Dirac distribution (see
formal definitions below).  Finally, the Green's function admits a
spectral expansion over the Steklov--Helmholtz eigenbasis:
\begin{equation}  \label{eq:Robin}
\widetilde G_q(\x,p|\x_0) = \widetilde{G}_D(\x,p|\x_0) + \frac{1}{D} \sum\limits_k \frac{V_k^{(p)}(\x) \, V_k^{(p)}(\x_0)}{\mu_k^{(p)} + q} \,,
\end{equation}
where $\widetilde{G}_D(\x,p|\x_0)$ is the Green's function of the
modified Helmholtz equation with mixed Dirichlet-Neumann conditions on
$\Gamma$ and $\Gamma_0$, respectively, and $\{\mu_k^{(p)},
V_k^{(p)}\}$ are the eigenvalues and eigenfunctions of the
Steklov-Helmholtz spectral problem \cite{Levitin}:
\begin{equation} \label{eq:Vk_def}
(p - D \Delta) V_k^{(p)} = 0 \quad (\x\in\Omega), \qquad \partial_n V_k^{(p)} = \mu_k^{(p)} V_k^{(p)} \quad (\x\in\Gamma),
\qquad \partial_n V_k^{(p)} = 0 \quad (\x\in \Gamma_0).
\end{equation}
The traces of $V_k^{(p)}$ on $\Gamma$ are the eigenfunctions of the
associated Dirichlet-to-Neumann (DtN) operator $\cM_p$.  The
elementary inverse Laplace transform of Eq.~\eqref{eq:Robin} with
respect to $q$ captures the explicit dependence on $\ell$,
\begin{equation}  \label{eq:Ptilde1_expansion}
\widetilde P(\x,\ell,p|\x_0) = \widetilde G_D(\x,p|\x_0) \delta(\ell) + \frac{1}{D}
\sum\limits_{k} V_k^{(p)}(\x) \,  V_k^{(p)}(\x_0) \, e^{-\mu_k^{(p)} \ell} \,,
\end{equation}
thus explaining why the Steklov--Helmholtz eigenbasis is so effective
for the single-target encounter problem.

For $N$ target regions $\Gamma_1, \ldots, \Gamma_N \subset \pa$, the
natural stochastic variables are the boundary local times
$\{\ell_{1,t}, \ldots, \ell_{N,t}\}$, accumulated on each target.  The
fundamental object thus becomes the encounter propagator
$P(\x,\ell_1,\ldots,\ell_N,t|\x_0)$.  Resolving encounters with
multiple targets {\it separately} provides access to their
correlations and competition and, more generally, allows distinct
reaction mechanisms to be imposed on each target.  However, no
analogue of the spectral expansion (\ref{eq:Ptilde1_expansion}) is
available for the general multi-target setting, except for a few
elementary geometries \cite{Grebenkov2020}.  The purpose of this work
is to develop such a generic-domain formulation.  The central
difficulty is not the construction of a new bulk eigenproblem.  The
ordinary DtN operator $\cM_p$ on the disconnected target set $\Gamma =
\Gamma_1\cup\cdots \cup \Gamma_N$ remains the natural operator, but
the $N$ independent boundary local times introduce target projection
operators that do not generally commute with $\cM_p$.  We show that a
decomposition into target-restricted blocks resolves this difficulty
and leads, after multi-dimensional Laplace inversion, to a convergent
switching expansion organized by successive transfers between the
targets.

The paper is organized as follows.  In Sec. \ref{sec:expansion}, we
derive the block-DtN representation, the switching expansion, its
target-restricted spectral form, and the singular decomposition
according to which targets have been encountered.  Section
\ref{sec:resummation} presents an exact operator-valued renewal
formulation and shows how resummation of the switching expansion in
the two-target setting naturally leads to modified Bessel functions
that emerge whenever the inter-target coupling preserves spectral
modes; it also discusses low-rank approximations for generic
geometries.  Section \ref{sec:small_targets} focuses on small,
well-separated targets and relates the resulting effective two-mode
description to matched-asymptotic encounter theory.  We finally
discuss extensions and open problems in Sec. \ref{sec:discussion}.

\section{Spectral expansion}
\label{sec:expansion}

In this section, we develop a generic-domain formulation of the
encounter propagator.  For clarity of presentation, we focus on the
two-target setting and then briefly describe its extension to an
arbitrary number of targets.  Throughout the manuscript, $D > 0$ is a
constant diffusion coefficient, which can be set to $1$ in all
equations (we retain it to keep physical units explicit).

\begin{figure}
\begin{center}
\includegraphics[width=42mm]{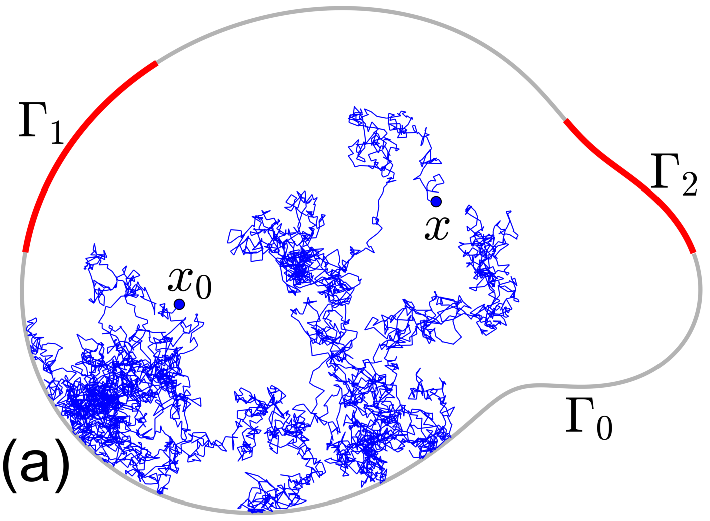} % ./simu/scheme_00.eps}
\includegraphics[width=42mm]{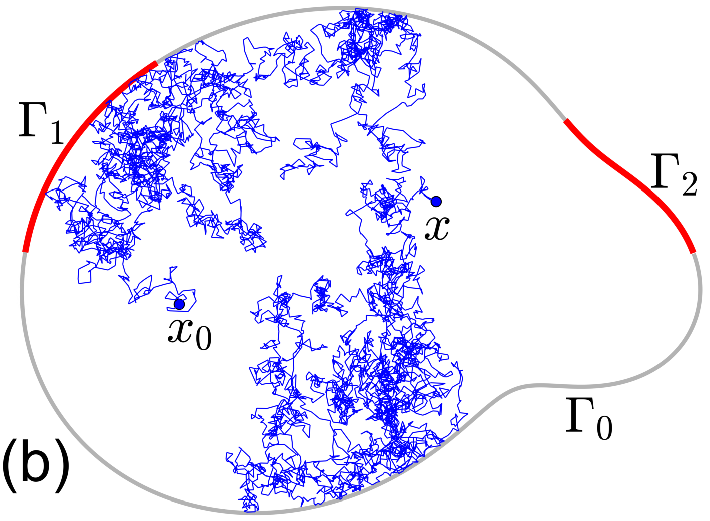} % ./simu/scheme_10.eps}
\includegraphics[width=42mm]{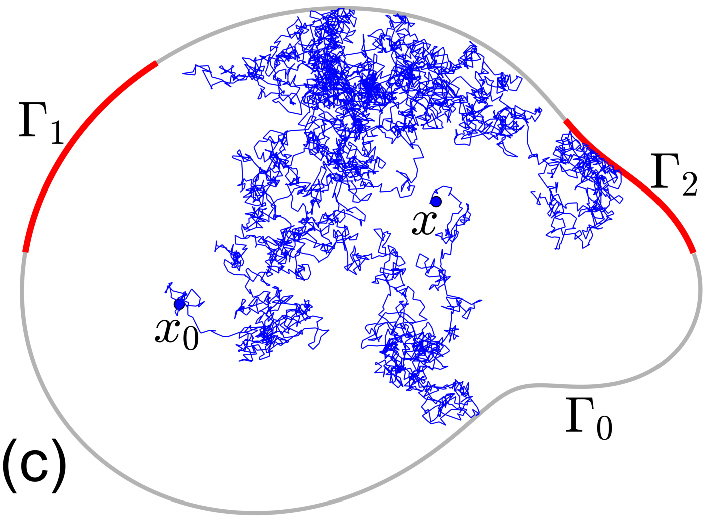} % ./simu/scheme_01.eps}
\includegraphics[width=42mm]{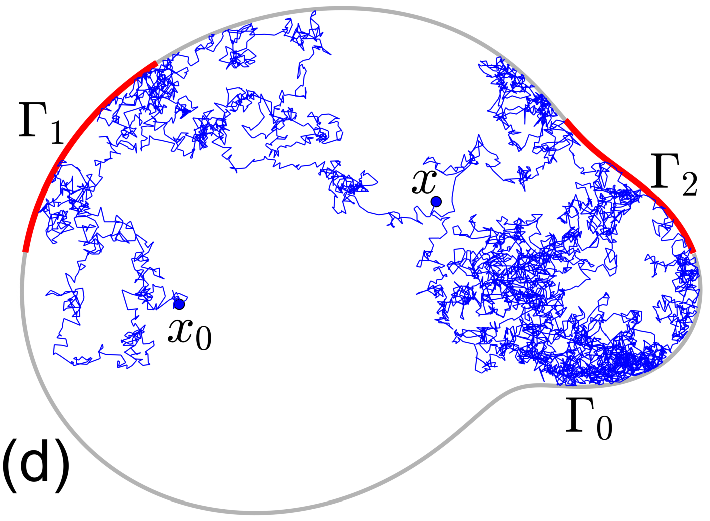} % ./simu/scheme_11.eps}
\end{center}
\caption{
Schematic representation of the two-target encounter problem in a
planar bounded domain $\Omega$, whose boundary $\pa$ is partitioned
into the reflecting part $\Gamma_0$ and two target regions $\Gamma_1$
and $\Gamma_2$.  During a given time $t$, the particle diffuses from a
point $\x_0$ to a point $\x$, and accumulates boundary local times
$\ell_{1,t}$ and $\ell_{2,t}$ on the target regions $\Gamma_1$ and
$\Gamma_2$.  Four representative cases are shown: (a) $\ell_{1,t} =
0$, $\ell_{2,t} = 0$; (b) $\ell_{1,t} > 0$, $\ell_{2,t} = 0$; (c)
$\ell_{1,t} = 0$, $\ell_{2,t} > 0$; and (d) $\ell_{1,t} > 0$,
$\ell_{2,t} > 0$.  Blue curve shows a simulated random trajectory in
each case.}
\label{fig:scheme}
% A_localtime9_two_scheme1b(ftype);
%
% load('traj_00d.mat');  % (a)
% load('traj_10.mat');   % (b)
% load('traj_01b.mat');   % (c)
% load('traj_11d.mat');  % (d)
% [X1,X2] = A_localtime9_two_scheme1(X1,X2);
\end{figure}

\subsection{Encounter propagator}

Let $\Omega\subset\mathbb{R}^d$ be a bounded connected domain with a
smooth boundary partitioned into three non-overlapping subsets,
$\partial\Omega = \Gamma_0\cup\Gamma_1\cup\Gamma_2$, where $\Gamma_0$
is reflecting and $\Gamma_1,\Gamma_2$ are two disjoint target regions
of smooth but arbitrary shape (Fig. \ref{fig:scheme}).  In particular,
each $\Gamma_i$ can be the union of a finite number of disconnected
subsets, whereas the boundary itself $\pa$ is not necessarily
connected so that $\Gamma_i$ can represent both boundary targets or
reactive surfaces located in the bulk.  We assume that
$\operatorname{dist}(\Gamma_1,\Gamma_2)>0$ to avoid possibly subtle
analysis at the junction between $\Gamma_1$ and $\Gamma_2$.  The
double Laplace transform of the encounter propagator
$P(\x,\ell_1,\ell_2,t|\x_0)$ with respect to two boundary local times
gives the usual heat kernel
\begin{align}
G_{q_1,q_2}(\x,t|\x_0) = & \int_0^\infty d\ell_1 \int_0^\infty d\ell_2\, e^{-q_1\ell_1-q_2\ell_2}
P(\x,\ell_1,\ell_2,t|\x_0),
\label{eq:double-transform-localtime}
\end{align}
which satisfies the diffusion equation,
\begin{equation}
\partial_tG_{q_1,q_2} = D \Delta G_{q_1,q_2} ,
\end{equation}
with the initial condition $G_{q_1,q_2}(\x,0|\x_0) = \delta(\x-\x_0)$
fixing the starting point $\x_0$, and mixed boundary conditions:
\begin{subequations}
\begin{align}
\partial_n G_{q_1,q_2} & =0, \quad \x\in\Gamma_0, \\
\partial_n G_{q_1,q_2} + q_1 G_{q_1,q_2} & =0, \quad \x\in\Gamma_1, \\
\partial_n G_{q_1,q_2} + q_2 G_{q_1,q_2} & =0, \quad \x\in\Gamma_2.
\end{align}
\end{subequations}
In turn, Laplace transformation in time gives the Green's function,
\begin{equation}
\widetilde G_{q_1,q_2}(\x,p|\x_0) = \int_0^\infty dt \, e^{-pt}\, G_{q_1,q_2}(\x,t|\x_0),
\end{equation}
which satisfies the modified Helmholtz equation
\begin{equation}
(p-D\Delta) \widetilde G_{q_1,q_2} = \delta(\x-\x_0),
\end{equation}
with the same boundary conditions.
The Laplace-domain encounter propagator is recovered through
\begin{align}
\widetilde P(\x,\ell_1,\ell_2,p|\x_0) = & \cL^{-1}_{q_1\to\ell_1} \cL^{-1}_{q_2\to\ell_2}
\left\{ \widetilde G_{q_1,q_2}(\x,p|\x_0) \right\}.
\label{eq:P-double-inversion}
\end{align}
A final inversion in $p$ yields the time-domain encounter propagator
$P(\x,\ell_1,\ell_2,t|\x_0)$.
The diagram
\begin{equation} \label{eq:diagram}
\begin{tikzcd}[column sep=huge, row sep=huge]
G_{q_1,q_2}(\x,t|\x_0)
 \arrow[r, shift left=1.2ex,
   "{\mathcal L^{-1}_{q_1\to\ell_1}
      \mathcal L^{-1}_{q_2\to\ell_2}}"]
 \arrow[d, shift left=1.2ex, "{\mathcal L_{t\to p}}"]
&
P(\x,\ell_1,\ell_2,t|\x_0)
 \arrow[l, shift left=1.2ex,
   "{\mathcal L_{\ell_1\to q_1}
      \mathcal L_{\ell_2\to q_2}}"]
 \arrow[d, shift left=1.2ex, "{\mathcal L_{t\to p}}"]
\\
\widetilde G_{q_1,q_2}(\x,p|\x_0)
 \arrow[u, shift left=1.2ex, "{\mathcal L^{-1}_{p\to t}}"]
 \arrow[r, shift left=1.2ex,
   "{\mathcal L^{-1}_{q_1\to\ell_1}
      \mathcal L^{-1}_{q_2\to\ell_2}}"]
&
\widetilde P(\x,\ell_1,\ell_2,p|\x_0)
 \arrow[l, shift left=1.2ex,
   "{\mathcal L_{\ell_1\to q_1}
      \mathcal L_{\ell_2\to q_2}}"]
 \arrow[u, shift left=1.2ex, "{\mathcal L^{-1}_{p\to t}}"]
\end{tikzcd}
\end{equation}
summarizes the direct and inverse Laplace transforms connecting the
encounter propagator to the heat kernel and the Green's function.

In principle, the encounter propagator $P$ could be recovered from
either $G_{q_1,q_2}$, or $\widetilde G_{q_1,q_2}$.  In practice,
however, the double Laplace inversion with respect to $q_1$ and $q_2$
is cumbersome both analytically and numerically.  In particular, an
inversion via the Bromwich integral would require resolving
$G_{q_1,q_2}$ or $\widetilde G_{q_1,q_2}$ for complex $q_1$ and $q_2$.
Our aim is instead to perform the local-time inversions analytically
at the operator level, leaving only the final inversion of $\widetilde
P$ with respect to $p$.  In other words, focusing on the bottom part
of the diagram~\eqref{eq:diagram}, we aim at finding the {\it
explicit} dependence of $\widetilde P(\x,\ell_1,\ell_2,p|\x_0)$ on
boundary local times $\ell_1$ and $\ell_2$, thus generalizing the
single-target expansion~\eqref{eq:Ptilde1_expansion} to two targets.
In this light, the central object of our study is not the encounter
propagator $P(\x,\ell_1,\ell_2,t|\x_0)$ itself, but its Laplace
transform $\widetilde P(\x,\ell_1,\ell_2,p|\x_0)$, which has its own
probabilistic interpretation and gives access to numerous
characteristics of the two-target diffusive search even without
Laplace inversion \cite{Grebenkov2020}.  In fact, introducing an
exponentially distributed random stopping time $\delta$ with the rate
$p$ and writing
\begin{equation}
p \widetilde P(\x,\ell_1,\ell_2,p|\x_0) = \int\limits_0^\infty dt \, \underbrace{p\, e^{-pt}}_{\textrm{PDF of}~\delta} \, P(\x,\ell_1,\ell_2,t|\x_0),
\end{equation}
one can interpret the left-hand side as the joint probability density
of the triple $\{ \bm{X}_\delta, \ell_{1,\delta}, \ell_{2,\delta}\}$,
i.e., the position and two boundary local times, evaluated at the
exponential stopping time $\delta$.  In some applications, the
stopping time $\delta$ can model the random lifetime of the diffusing
species \cite{Yuste13,Meerson15,Grebenkov17} so that $p \widetilde
P(\x,\ell_1,\ell_2,p|\x_0)$ fully describes the position of such a
mortal particle and its interactions with both targets until death.

Finally, we note that the goal of this study can also be formulated as
a practical problem from applied mathematics: how to evaluate
explicitly the double inverse Laplace transform of the Green's
function $\widetilde G_{q_1,q_2}$ with respect to the Robin parameters
$q_1$ and $q_2$?

\subsection{Dirichlet-to-Neumann operator on the union of targets}
\label{sec:DtN_def}

For fixed $p\geq 0$ and given functions $f_i \in
H^{\frac12}(\Gamma_i)$, let $u$ solve the boundary value problem
\begin{equation}  \label{eq:u_Helm_eq}
(p-D\Delta)u=0  \qquad\text{in }\Omega,
\qquad u|_{\Gamma_1}=f_1, \qquad    u|_{\Gamma_2}=f_2,    \qquad   \partial_n u|_{\Gamma_0}=0.
\end{equation}
The DtN operator is defined by
\begin{equation}
\cM_p
\begin{pmatrix}
f_1\\
f_2
\end{pmatrix}
=
\begin{pmatrix}
\partial_nu|_{\Gamma_1}\\
\partial_nu|_{\Gamma_2}
\end{pmatrix},
\label{eq:DtN-def}
\end{equation}
and maps the Dirichlet trace space into its dual.  We subsequently
regard $\mathcal M_p$ as a self-adjoint operator in the Hilbert space
\begin{equation*}
\cH = L^2(\Gamma_1)\oplus L^2(\Gamma_2) \simeq L^2(\Gamma), \qquad \Gamma = \Gamma_1\cup \Gamma_2.  
\end{equation*}
Under the above regularity assumptions, the self-adjoint realization
of $\mathcal M_p$ in $\mathcal H$ has compact resolvent and hence a
discrete spectrum \cite{Levitin}, with a countable set of eigenpairs
$\{\mu_n^{(p)}, v_n^{(p)}\}$, satisfying
\begin{equation}
\cM_pv_n^{(p)} = \mu_n^{(p)} v_n^{(p)}.
\end{equation}

Let $P_1$ and $P_2$ denote the orthogonal projections onto the two
target spaces such that $P_1+P_2=I$.  Consequently, the multiparameter
DtN resolvent is
\begin{equation}
\cR_p(q_1,q_2) = \left(\cM_p + q_1P_1 + q_2P_2\right)^{-1}.
\label{eq:fundamental-resolvent}
\end{equation}
If $q_1 = q_2 = q$, one has $q_1 P_1 + q_2 P_2 = q I$, and the
eigenbasis of the DtN operator $\cM_p$ diagonalizes this resolvent,
resulting in the spectral expansion~\eqref{eq:Ptilde1_expansion}.
This special case corresponds to probing only the total boundary local
time $\ell=\ell_1+\ell_2$, i.e., to treating $\Gamma = \Gamma_1 \cup
\Gamma_2$ as a single target.  However, in a generic situation $q_1
\ne q_2$, the matrix representation of the resolvent in this
eigenbasis is
\begin{equation*}
\cR_p(q_1,q_2) \longleftrightarrow \left[\Lambda_p+q_1A_p+q_2(I-A_p) \right]^{-1},
\end{equation*}
where $(\Lambda_p)_{nm} = \mu_n^{(p)} \delta_{nm}$ is the diagonal
(infinite-dimensional) matrix representing $\cM_p$, and
\begin{equation*} 
(A_p)_{nm} = \langle v_n^{(p)},P_1v_m^{(p)}\rangle
= \int_{\Gamma_1} [v_n^{(p)}(\bm{s})]^* v_m^{(p)}(\bm{s})\,d\bm{s}.
\end{equation*}
This representation shows the basic obstruction in a generic geometry:
$[\cM_p,P_1]\neq 0$.  Hence the ordinary DtN eigenbasis does not
simultaneously diagonalize the dependence on $q_1$ and $q_2$, thus
prohibiting an explicit inversion of the double Laplace transform via
an analogue of~\eqref{eq:Ptilde1_expansion}.

Throughout the following analysis we take $p > 0$, while $p = 0$ is
treated as a limit.  The corresponding DtN operator is therefore real
and self-adjoint, and its eigenfunctions may be chosen real.  We
nevertheless retain complex conjugation in the formulas, as this
notation remains applicable to a possible analytic continuation in
$p$.  Such a continuation, which is required, e.g., for a Bromwich
inversion, involves a non-self-adjoint extension of the present
construction and will not be considered here.

\subsection{Robin Green's function representation}

Let $\widetilde G_D(\x,p|\x_0)$ denote the Green's function with
Dirichlet conditions on both targets and Neumann condition on
$\Gamma_0$:
\begin{subequations}
\begin{align}
(p-D\Delta) \widetilde G_D(\x,p|\x_0) &= \delta(\x-\x_0), \quad \x\in \Omega,\\
\widetilde G_D&=0, \quad\x\in\Gamma_1\cup\Gamma_2, \\
\partial_n \widetilde G_D&=0, \quad \x\in\Gamma_0.
\end{align}
\end{subequations}
Define the boundary flux vector
\begin{equation}
\bm{j}_p(\x) =
\begin{pmatrix}
j_{1,p}(\cdot|\x)\\
j_{2,p}(\cdot|\x)
\end{pmatrix},
\qquad \textrm{with}\quad
j_{i,p}(\bm{s}|\x) = -D \partial_{n_{\bm{s}}} \widetilde G_D(\bm{s},p|\x), \quad \bm{s}\in\Gamma_i.
\end{equation}
The Robin Green's function then admits the representation
\begin{equation}
\widetilde G_{q_1,q_2}(\x,p|\x_0) = \widetilde G_D(\x,p|\x_0) + \frac{1}{D}
\left\langle \bm{j}_p(\x), \cR_p(q_1,q_2) \bm{j}_p(\x_0) \right\rangle_{\cH},
\label{eq:Green-resolvent}
\end{equation}
where the inner product includes the sum over the two target
components.  Defining the operator-valued encounter kernel
\begin{equation}
\cK_p(\ell_1,\ell_2) = \cL^{-1}_{q_1\to\ell_1} \cL^{-1}_{q_2\to\ell_2} \cR_p(q_1,q_2),
\label{eq:K-def}
\end{equation}
one has
\begin{equation}
\widetilde P(\x,\ell_1,\ell_2,p|\x_0) = \widetilde G_D(\x,p|\x_0) \delta(\ell_1)\delta(\ell_2)
+ \frac{1}{D} \left\langle \bm{j}_p(\x), \cK_p(\ell_1,\ell_2) \bm{j}_p(\x_0) \right\rangle_{\cH}.
\label{eq:P-master}
\end{equation}
This is a precursor for generalizing the single-target spectral
expansion~\eqref{eq:Ptilde1_expansion}.  The generic two-target
problem thus reduces to finding $\cK_p(\ell_1,\ell_2)$.

\subsection{Target-block representation}

Decompose the DtN operator as
\begin{equation}
\cM_p = \begin{pmatrix}
M_{11}^{(p)}&M_{12}^{(p)}\\
M_{21}^{(p)}&M_{22}^{(p)}
\end{pmatrix} = 
\begin{pmatrix}
A & C\\
C^\dagger& B
\end{pmatrix},
\end{equation}
where we set for notational simplicity:
\begin{equation}
A=M_{11}^{(p)} , \qquad B=M_{22}^{(p)}, \qquad C=M_{12}^{(p)}, \qquad C^\dagger=M_{21}^{(p)}.
\end{equation}
Then
\begin{equation}
\cR_p(q_1,q_2) =(D_\q+V)^{-1} ,
\label{eq:block-resolvent}
\qquad \textrm{where} \quad
D_\q= \begin{pmatrix}
A+q_1I&0\\
0&B+q_2I
\end{pmatrix},
\qquad
V=
\begin{pmatrix}
0&C\\
C^\dagger&0
\end{pmatrix}, 
\end{equation}
with $q = (q_1,q_2)$.  For sufficiently large
$\operatorname{Re}\{q_i\}$, the resolvent admits the Neumann expansion
\begin{equation}
\cR_p = D_\q^{-1} - D_\q^{-1}VD_\q^{-1} + D_\q^{-1}VD_\q^{-1}VD_\q^{-1} -\cdots.
\label{eq:Neumann}
\end{equation}
Introducing
\begin{equation}
R_1(q_1)=(A+q_1I)^{-1}, \qquad R_2(q_2)=(B+q_2I)^{-1},
\end{equation}
and using the elementary Laplace inversions
$\cL^{-1}_{q_1\to\ell_1} \{R_1(q_1)\} = e^{-A\ell_1}$ and
$\cL^{-1}_{q_2\to\ell_2} \{R_2(q_2)\} = e^{-B\ell_2}$,
one can invert Eq.~\eqref{eq:Neumann} term by term.

\begin{figure}
\begin{center}
\includegraphics[width=50mm]{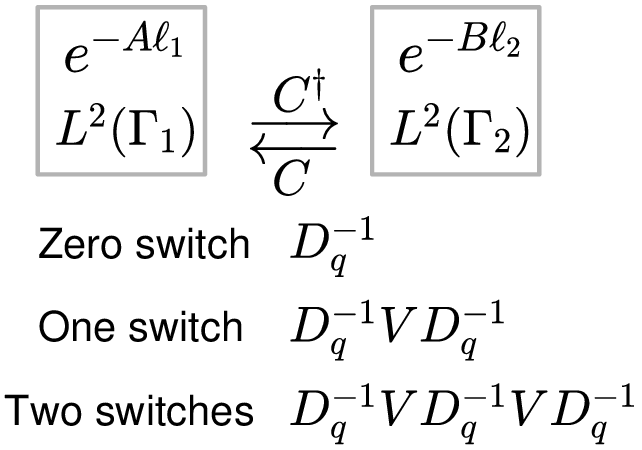} % ./simu/scheme2.eps}
\end{center}
\caption{
In the target-block representation, the diagonal DtN operators $A$ and
$B$ generate evolution in the corresponding local-time variables,
whereas the off-diagonal blocks $C$ and $C^\dagger$ transfer boundary
data between the two target spaces.  Successive terms of the switching
expansion are therefore associated with alternating sequences of
target labels.  The arrows represent the operator structure of the
expansion and should not be interpreted as a unique decomposition of
individual Brownian trajectories.  }
\label{fig:switching}
% A_localtime9_two_scheme2;
\end{figure}

The block-diagonal resolvents $R_1$ and $R_2$ in $D_\q^{-1}$ describe
local-time evolution associated with each target separately, whereas
each factor $C$ or $C^\dagger$ in $V$ transfers the boundary data from
one target space to the other (Fig. \ref{fig:switching}).  This
motivates interpreting successive powers of $V$ as successive
target-to-target switches.  As shown below, this interpretation
becomes exact at the level of the inverse Laplace transform and does
not rely on a weak-coupling assumption.

\subsection{Switching expansion of the encounter kernel}

Write
\begin{equation}
\cK_p(\ell_1,\ell_2) =
\begin{pmatrix}
K_{11}&K_{12}\\
K_{21}&K_{22}
\end{pmatrix} ,
\end{equation}
where each block $K_{ij}$ can be obtained from the double Laplace
inversion of the corresponding block of $\cR_p$ given by the Neumann
expansion~\eqref{eq:Neumann}.

Let us start with the diagonal block $K_{11}$.  Writing the first term
in Eq.~\eqref{eq:Neumann} as
\begin{equation*}
D_\q^{-1} = \begin{pmatrix}
R_1(q_1) & 0 \\
0& R_2(q_2)
\end{pmatrix} ,
\end{equation*}
one obtains immediately the zero-switch contribution to the diagonal
block $K_{11}$:
\begin{equation*}
K_{11}^{(0)}(\ell_1,\ell_2) = e^{-A\ell_1}\delta(\ell_2).
\label{eq:K110}
\end{equation*}
The second term in Eq.~\eqref{eq:Neumann} represents one switch
between two targets,
\begin{equation*}
D_\q^{-1} V D_\q^{-1} = \begin{pmatrix}
0 & R_1(q_1) C R_2(q_2) \\
R_2(q_2) C^\dagger R_1(q_1) & 0 
\end{pmatrix} ,
\end{equation*}
which does not contribute to the diagonal block $K_{11}$.  In turn,
the third term accounts for two switches and has the form
\begin{equation*}
D_\q^{-1} V D_\q^{-1} V D_\q^{-1} = \begin{pmatrix}
R_1(q_1) C R_2(q_2) C^\dagger R_1(q_1) & 0  \\
0& R_2(q_2) C^\dagger R_1(q_1) C R_2(q_2) 
\end{pmatrix} .
\end{equation*}
The first block corresponds to inter-target transfers,
$\Gamma_1\to\Gamma_2\to\Gamma_1$, and its double inverse Laplace
transform reads
\begin{align*}
K_{11}^{(2)}(\ell_1,\ell_2) =& \int_0^{\ell_1}d\ell_1'\, e^{-A\ell_1'} C e^{-B\ell_2} C^\dagger e^{-A(\ell_1-\ell_1')}.
\end{align*}
We rewrite this integral formally as
\begin{align*}
K_{11}^{(2)}(\ell_1,\ell_2) =& \int_{\ell_1^1+\ell_1^2=\ell,~ \ell_1^j \geq0} d\ell_1^1 d\ell_1^2\,
 e^{-A\ell_1^1} C e^{-B\ell_2} C^\dagger e^{-A\ell_1^2}.
\end{align*}

More generally, the contribution of $2m$ switches between targets is
\begin{align}
K_{11}^{(2m)}(\ell_1,\ell_2)
={}&
\int_{\substack{\ell_1^1+\cdots+\ell_1^{m+1}=\ell_1\\ \ell_1^j\ge0}} d\ell_1^1 \cdots d\ell_1^{m+1}
\int_{\substack{\ell_2^1+\cdots+\ell_2^m=\ell_2\\ \ell_2^j\ge0}} d\ell_2^1 \cdots d\ell_2^{m}
\nonumber\\
&\times
e^{-A\ell_1^1} C e^{-B\ell_2^1}C^\dagger e^{-A\ell_1^2} \cdots C e^{-B\ell_2^m}C^\dagger e^{-A\ell_1^{m+1}},
\label{eq:K11-general}
\end{align}
so that
\begin{equation}
K_{11} = \delta(\ell_2) e^{-A\ell_1} + \sum_{m=1}^{\infty} K_{11}^{(2m)}.
\label{eq:K11-series}
\end{equation}
The same computation yields 
\begin{equation}
K_{22} = \delta(\ell_1)e^{-B\ell_2} + \sum_{m=1}^{\infty}K_{22}^{(2m)},
\label{eq:K22-series}
\end{equation}
where $K_{22}^{(2m)}$ is given by Eq.~\eqref{eq:K11-general}, in which
$A$ is exchanged with $B$, and $\ell_1$ is exchanged with $\ell_2$.
These expansions have a direct encounter interpretation: each factor $C$
or $C^\dagger$ switches the target on which the boundary local time is
subsequently accumulated, while the semigroups $e^{-A\ell_1}$ and
$e^{-B\ell_2}$ describe evolution in the corresponding local-time
variable between successive switches.

For off-diagonal blocks, an odd number of transfers is required to
connect the two target spaces.  The lowest-order contribution to the
off-diagonal block $K_{12}$ is
\begin{equation*}
K_{12}^{(1)}(\ell_1,\ell_2) = - e^{-A\ell_1}C e^{-B\ell_2}.
\end{equation*}
The general term with $2m+1$ transfers is
\begin{align}
K_{12}^{(2m+1)} =& - \int_{\substack{\ell_1^1+\cdots+\ell_1^{m+1}=\ell_1\\ \ell_1^j\ge0}} d\ell_1^1 \cdots d\ell_1^{m+1}
\int_{\substack{\ell_2^1+\cdots+\ell_2^{m+1}=\ell_2\\ \ell_2^j\ge0}} d\ell_2^1 \cdots d\ell_2^{m+1}
\nonumber\\
&\times
e^{-A\ell_1^1}C e^{-B\ell_2^1} C^\dagger e^{-A\ell_1^2} C e^{-B\ell_2^2} \cdots C e^{-B\ell_2^{m+1}},
\label{eq:K12-general}
\end{align}
so that
\begin{equation}  \label{eq:K12-series}
K_{12} = \sum_{m=0}^{\infty} K_{12}^{(2m+1)}.
\end{equation}
The other block follows by Hermitian conjugation,
\begin{equation}
K_{21}(\ell_1,\ell_2) = K_{12}(\ell_1,\ell_2)^\dagger ,
\end{equation}
where the adjoint naturally exchanges the source and target boundary
spaces.

For $p>0$, the diagonal DtN blocks are non-negative self-adjoint
operators, so that
\begin{equation}
\|e^{-A\ell_1}\|\le1, \qquad \|e^{-B\ell_2}\|\le1.
\end{equation}
For separated targets, $C$ is bounded.  The simplex volumes in
Eq.~\eqref{eq:K12-general} yield
\begin{equation}
\left\| K_{12}^{(2m+1)} \right\| \le \|C\|^{2m+1} \frac{\ell_1^m}{m!} \frac{\ell_2^m}{m!} \,.
\label{eq:K12-bound}
\end{equation}
Similarly, for $m\ge1$,
\begin{equation}
\left\| K_{11}^{(2m)} \right\| \le \|C\|^{2m} \frac{\ell_1^m}{m!} \frac{\ell_2^{m-1}}{(m-1)!} \,.
\label{eq:K11-bound}
\end{equation}
The switching series in~(\ref{eq:K11-series}, \ref{eq:K22-series},
\ref{eq:K12-series}) therefore converge absolutely in operator norm
for every finite $\ell_1,\ell_2$ under these assumptions.

\subsection{Target-restricted DtN eigenbasis}

The block representation suggests diagonalizing $A$ and $B$
separately:
\begin{equation}  \label{eq:A-spectrum}
A\phi_n^{(p)} = \alpha_n^{(p)} \phi_n^{(p)},  \qquad 
B\psi_m^{(p)} = \beta_m^{(p)} \psi_m^{(p)}.
\end{equation}
In bulk form, the first eigenvalue problem is equivalent to the
Steklov--Helmholtz spectral problem
\begin{subequations}
\begin{align}
(p-D\Delta) \Phi_n^{(p)}& = 0, \quad \x\in\Omega, \\
\partial_n \Phi_n^{(p)} &= \alpha_n^{(p)} \Phi_n^{(p)}, \quad \x\in\Gamma_1, \\  \label{eq:Phi_Dirichlet}
\Phi_n^{(p)} &=0, \quad \x\in\Gamma_2, \\
\partial_n \Phi_n^{(p)} &=0, \quad \x\in\Gamma_0,
\end{align}
\end{subequations}
so that the trace of $\Phi_n^{(p)}$ on $\Gamma_1$ is $\phi_n^{(p)}$.
The key difference from the global Steklov--Helmholtz
problem~\eqref{eq:Vk_def} on $\Gamma = \Gamma_1 \cup \Gamma_2$ is that
the second target $\Gamma_2$ is now ``excluded'' by setting the
Dirichlet condition~\eqref{eq:Phi_Dirichlet}.  The bulk problem for
$\psi_m^{(p)}$ is similar, with exchange of $\Gamma_1$ and $\Gamma_2$
and substitution of $\alpha_n^{(p)}$ by $\beta_n^{(p)}$.

The most convenient basis for the joint encounter problem is thus
$\{\phi_n^{(p)}\}\oplus\{\psi_m^{(p)}\}$, together with the
inter-target coupling coefficients
\begin{equation}
c_{nm}(p) = \langle \phi_n^{(p)},C \psi_m^{(p)} \rangle_{\cH} = \int\limits_{\Gamma_1} d\bm{s} \, [\phi_n^{(p)}]^* \, C \psi_m^{(p)},
\label{eq:coupling}
\end{equation}
given that the trace of $\Phi_n^{(p)}$ on $\Gamma_2$ is zero.  In the
same way, one has
\begin{equation}
d_{mn}(p) = \int\limits_{\Gamma_2} d\bm{s} \, [\psi_m^{(p)}]^* \, C^\dagger \phi_n^{(p)} 
=  \langle \psi_m^{(p)},C^\dagger \phi_n^{(p)} \rangle_{\cH}
= \langle C \psi_m^{(p)}, \phi_n^{(p)} \rangle_{\cH} = c_{nm}^*(p),
\end{equation}
because $C^\dagger$ is the adjoint of $C$.  In the following, we use
$c_{nm}(p)$ for representing $C$ and $c_{nm}^*(p)$ for representing
$C^\dagger$.  For smooth, positively separated targets, the smoothing
character of $C$ implies decay of its matrix elements in regular
spectral bases, suggesting efficient spectral truncation.

These eigenbases yield explicit spectral representations of the
switching terms.  The simplest off-diagonal term becomes
\begin{equation*}
K_{12}^{(1)} = - \sum_{n,m} c_{nm}(p) e^{-\alpha_n^{(p)}\ell_1 - \beta_m^{(p)}\ell_2}
|\phi_n^{(p)}\rangle \langle\psi_m^{(p)}|.
\end{equation*}
To get spectral representations of higher-order terms, we introduce
the simplex exponential integral
\begin{align}
\Phi_m(\ell; a_1,\ldots,a_m) = & \int_{\substack{\ell^1+\cdots+\ell^m=\ell\\ \ell^j\ge0}}
\exp\left(-\sum_{j=1}^m a_j \ell^j \right) d\ell^1 \cdots d\ell^m , \qquad a_1,\ldots,a_m \geq 0.
\label{eq:Phi-def}
\end{align}
For pairwise distinct $a_j$, one has
\begin{equation}
\Phi_m(\ell;a_1,\ldots,a_m) = \sum_{j=1}^{m} \frac{e^{-a_j\ell}} {\displaystyle\prod_{k\ne j}(a_k-a_j)}
\label{eq:Phi-explicit}
\end{equation}
(when some $a_j$ are identical, $\Phi_m$ can be evaluated by
continuity, yielding polynomial factors multiplying exponentials).

Using Eqs.~\eqref{eq:A-spectrum} and~\eqref{eq:coupling}, the
$2m$-switch contribution to $K_{11}$ becomes
\begin{align}
K_{11}^{(2m)} =& \sum_{\substack{n_0,\ldots,n_m\\r_1,\ldots,r_m}}
\left[\prod_{k=1}^{m} c_{n_{k-1}r_k} c_{n_kr_k}^{*} \right]
\Phi_{m+1}(\ell_1; \alpha_{n_0},\ldots,\alpha_{n_m})
\Phi_{m}(\ell_2; \beta_{r_1},\ldots,\beta_{r_m})
|\phi_{n_0}\rangle \langle\phi_{n_m}|.
\label{eq:K11-spectral}
\end{align}
Likewise,
\begin{align}
K_{12}^{(2m+1)} = &- \sum_{\substack{n_0,\ldots,n_m\\r_0,\ldots,r_m}}
c_{n_0r_0} c_{n_1r_0}^{*} c_{n_1r_1} \cdots c_{n_mr_m}
\Phi_{m+1}(\ell_1; \alpha_{n_0},\ldots,\alpha_{n_m})
\Phi_{m+1}(\ell_2; \beta_{r_0},\ldots,\beta_{r_m})
|\phi_{n_0}\rangle \langle\psi_{r_m}|.
\label{eq:K12-spectral}
\end{align}
Equations~\eqref{eq:K11-spectral} and \eqref{eq:K12-spectral},
together with their target-exchanged counterparts, provide an explicit
generic-domain spectral representation of the operator-valued
encounter kernel.

\subsection{Singular structure of the encounter propagator}

Equation~\eqref{eq:P-master} naturally separates the encounter propagator
into four parts:
\begin{align}
\widetilde P(\x,\ell_1,\ell_2,p|\x_0) = & \widetilde P_{00}(\x,p|\x_0) \delta(\ell_1)\delta(\ell_2)
+ \widetilde P_{10}(\x,\ell_1,p|\x_0) \delta(\ell_2)
\nonumber\\
&+ \widetilde P_{01}(\x,\ell_2,p|\x_0) \delta(\ell_1)
+ \widetilde P_{11}(\x,\ell_1,\ell_2,p|\x_0).
\label{eq:P-decomposition}
\end{align}
This decomposition is distributional in $(\ell_1,\ell_2)$ and reflects
four mutually exclusive classes of encounter histories, as illustrated
in Fig. \ref{fig:scheme}:

(i) The first term describes trajectories that have not encountered
any target and thus accumulated no boundary local time on either
target:
\begin{equation}
\widetilde P_{00}(\x,p|\x_0) = \widetilde G_D(\x,p|\x_0).
\end{equation}
This is naturally represented by the Dirichlet Green's function that
penalizes any encounter with $\Gamma_1$ and $\Gamma_2$.

(ii) The second term describes contributions associated with
encounters with $\Gamma_1$ but not $\Gamma_2$:
\begin{equation}
\widetilde P_{10} = \frac{1}{D} \left\langle j_{1,p}(\x), e^{-A\ell_1} j_{1,p}(\x_0) \right\rangle_{\Gamma_1}, \qquad \ell_1>0.
\label{eq:P10}
\end{equation}

(iii) Similarly, the third term is
\begin{equation}
\widetilde P_{01} = \frac{1}{D} \left\langle j_{2,p}(\x), e^{-B\ell_2} j_{2,p}(\x_0) \right\rangle_{\Gamma_2}, \qquad \ell_2>0.
\label{eq:P01}
\end{equation}
Both terms have a simple interpretation: they are single-target
encounter propagators in which the other target is kept absorbing.

(iv) The remaining contribution
\begin{equation}
\widetilde P_{11}(x,\ell_1,\ell_2,p|\x_0), \qquad \ell_1>0,\quad\ell_2>0,
\end{equation}
contains all switching terms involving the off-diagonal DtN blocks,
and describes trajectories that have encountered both target regions.
In block form,
\begin{align}
\widetilde P_{11} = \frac{1}{D} \sum_{i,j=1}^{2} \left\langle j_{i,p}(\x), K_{ij,p}^{\rm reg}(\ell_1,\ell_2) j_{j,p}(\x_0)
\right\rangle_{\cH},
\label{eq:P11}
\end{align}
where $K_{ij,p}^{\rm reg}$ denotes the part of the corresponding block
supported in the open quadrant $\ell_1,\ell_2>0$ (i.e., by excluding
singular terms).
This decomposition is the generic-domain counterpart of the singular
structures that arise explicitly in elementary two-target geometries
\cite{Grebenkov2020}.

\subsection{Extension to an arbitrary number of targets}

Let $\partial\Omega=\bigcup_{i=0}^{N}\Gamma_i$ be a union of $N$
disjoint target regions $\Gamma_i$, complemented by the reflecting
part $\Gamma_0$, and let $P_i$ denote the projection operator onto
$L^2(\Gamma_i)$.  The multi-local-time transform of the encounter
propagator is governed by the boundary resolvent
\begin{equation}
\cR_p(q_1,\ldots,q_N) = \left(\cM_p+\sum_{i=1}^{N}q_iP_i \right)^{-1}.
\label{eq:N-resolvent}
\end{equation}
The operator-valued encounter kernel is
\begin{equation}
\cK_p(\ell_1,\ldots,\ell_N) =
\left(\prod_{i=1}^{N} \cL^{-1}_{q_i\to\ell_i} \right) \cR_p(q_1,\ldots,q_N).
\label{eq:N-kernel}
\end{equation}
A block decomposition of $\cM_p$ with respect to
$\bigoplus_{i=1}^{N}L^2(\Gamma_i)$ leads to the Neumann
series~\eqref{eq:Neumann}, which can again be inverted term by term.
Each term can then be associated with a walk on the complete graph of
target labels, with the off-diagonal DtN blocks providing the
transition operators.  The two-target switching series is thus the
simplest member of a general expansion in target-to-target
transitions.

\section{Resummation of the switching expansion}
\label{sec:resummation}

The switching expansion derived above provides an explicit convergent
representation of the encounter kernel for a generic geometry.  It is
natural to ask whether this infinite series can be resummed, as occurs
in elementary geometries for which the Laplace-transformed encounter
propagator can be expressed in terms of modified Bessel functions
\cite{Grebenkov2020}.  In this section, we show that an exact
resummation is always possible at the operator level, whereas its
reduction to scalar special functions relies essentially on the
absence of spectral mode mixing between the two targets.  This
observation also suggests intermediate approximations for generic
geometries.

\subsection{Operator-valued renewal representation}

Using the above notations, the first block of the resolvent
$\cR_p(q_1,q_2)$ admits the Schur-complement representation
\begin{equation}
R_{11} = \left[A+q_1I-C(B+q_2I)^{-1}C^\dagger \right]^{-1}
= \left[I-R_1(q_1) C R_2(q_2) C^\dagger \right]^{-1}R_1(q_1).
\label{eq:R11-geometric-resum}
\end{equation}
Equation~\eqref{eq:R11-geometric-resum} is an exact operator-level
resummation of all even switching contributions.  This equation can be
written in a renewal-type form in $(q_1,q_2)$ variables:
\begin{equation}
R_{11} = R_1(q_1) + R_1(q_1) C R_2(q_2) C^\dagger R_{11}.
\label{eq:R11-geometric-resum2}
\end{equation}

An equivalent representation can be obtained directly in the
local-time variables.  For this purpose, let us define the
two-dimensional convolution of operator-valued kernels $F$ and $G$ as
\begin{align}
(F*G)(\ell_1,\ell_2) = & \int_0^{\ell_1}d\ell_1' \int_0^{\ell_2}d\ell_2' \, F(\ell_1',\ell_2') \, G(\ell_1-\ell_1',\ell_2-\ell_2').
\label{eq:2d-convolution}
\end{align}
The double inverse Laplace transform
of~\eqref{eq:R11-geometric-resum2} with respect to $q_1$ and $q_2$ can
thus be written as the operator-valued renewal equation
\begin{equation}
K_{11} = e^{-A\ell_1} \delta(\ell_2) + T*K_{11},
\label{eq:K11-renewal}
\end{equation}
where 
\begin{equation}
T(\ell_1,\ell_2) = \cL^{-1}_{q_1\to\ell_1} \cL^{-1}_{q_2\to\ell_2} \bigl\{ R_1(q_1) C R_2(q_2) C^\dagger \bigr\} 
= e^{-A\ell_1} C e^{-B\ell_2}C^\dagger.
\label{eq:roundtrip-kernel}
\end{equation}
Equivalently, introducing the convolution operator $T*$, one may write
\begin{equation}
K_{11} = (I-T*)^{-1} e^{-A\ell_1} \delta(\ell_2).
\label{eq:K11-renewal-resummed}
\end{equation}
Expanding the inverse in Eq.~\eqref{eq:K11-renewal-resummed}
reproduces precisely the even-switch series (\ref{eq:K11-general},
\ref{eq:K11-series}) derived above.  A similar representation holds
for the diagonal block $K_{22}$:
\begin{equation}
K_{22} = e^{-B\ell_2} \delta(\ell_1) + \bigl(e^{-B\ell_2} C^\dagger e^{-A\ell_1}C\bigr) *K_{22}.
\label{eq:K22-renewal}
\end{equation}

The off-diagonal block can be reconstructed from $K_{11}$.  As the
Schur-complement representation of $R_{12}$ reads
\begin{equation}
R_{12} = - \left[I-R_1(q_1) C R_2(q_2) C^\dagger \right]^{-1}R_1(q_1) C R_2(q_2) = - R_{11} C R_2(q_2),
\end{equation}
its double Laplace inversion yields
\begin{equation}
K_{12} = -K_{11}* \bigl[C e^{-B\ell_2}\delta(\ell_1)\bigr].
\label{eq:K12-renewal}
\end{equation}
Alternatively, one can get the operator-valued renewal equation
\begin{equation}
K_{12} = - e^{-A\ell_1} C e^{-B\ell_2} + T * K_{12}.
\label{eq:K12-renewal2}
\end{equation}
A similar construction holds for $K_{21}$.  Thus, even in a completely
generic geometry, the infinite switching expansion admits an exact
resummation in terms of an operator-valued two-dimensional renewal
problem.

\subsection{Mode-preserving coupling}

The origin of the explicit modified-Bessel-function expressions found
in simple geometries becomes particularly transparent within the
present formalism.  
Suppose that the target-restricted DtN bases can be chosen such that
the inter-target coupling does not mix different modes.  The full DtN
operator then decomposes into mutually orthogonal two-dimensional
invariant subspaces,
\begin{equation*}
\cH = \bigoplus_\nu \cH_\nu, \qquad \cH_\nu = \mathrm{span}\{ (\phi_\nu,0),\, (0,\psi_\nu)\},
\end{equation*}
and its restriction to each subspace is represented by the $2\times2$
matrix
\begin{equation}  
\mathcal M_p \biggl|_{\cH_\nu} = \begin{pmatrix}
a_\nu & c_\nu\\
c_\nu^* & b_\nu
\end{pmatrix}.
\label{eq:2by2-decomposition}
\end{equation}
This means that
\begin{equation}
A\phi_\nu =a_\nu\phi_\nu,  \qquad  B\psi_\nu = b_\nu\psi_\nu,  \qquad C\psi_\nu =c_\nu \phi_\nu,
\qquad C^\dagger\phi_\nu=c_\nu^*\psi_\nu.
\label{eq:channel-diagonal}
\end{equation}

Suppressing the mode index $\nu$ for clarity, let $A=a$, $B=b$, and
$C=c$.  The off-diagonal switching series becomes
\begin{equation}
K_{12}(\ell_1,\ell_2) = - c\,e^{-a\ell_1-b\ell_2} \sum_{m=0}^{\infty} \frac{ \left(|c|^2\ell_1\ell_2\right)^m}{(m!)^2}
= -c\, e^{-a\ell_1-b\ell_2} I_0\left(2|c|\sqrt{\ell_1\ell_2}\right),
\label{eq:K12-Bessel}
\end{equation}
where we identified the series expansion of the modified Bessel
function $I_0(z)$ of the first kind.  Similarly,
\begin{equation}
K_{21}(\ell_1,\ell_2) = -c^*\, e^{-a\ell_1-b\ell_2} I_0\left(2|c|\sqrt{\ell_1\ell_2} \right).
\label{eq:K21-Bessel}
\end{equation}

For the diagonal block, the regular part of the switching series reads
\begin{equation*}
K_{11}^{\rm reg}(\ell_1,\ell_2) = e^{-a\ell_1-b\ell_2} \sum_{m=1}^{\infty}
\frac{|c|^{2m}\ell_1^m\ell_2^{m-1}}{m!(m-1)!} \,,
\end{equation*}
so that
\begin{equation}
K_{11}(\ell_1,\ell_2) = e^{-a\ell_1} \delta(\ell_2)
+ |c| \sqrt{\frac{\ell_1}{\ell_2}}\, e^{-a\ell_1-b\ell_2} I_1\left(2|c|\sqrt{\ell_1\ell_2}\right).
\label{eq:K11-Bessel}
\end{equation}
Likewise,
\begin{equation}
K_{22}(\ell_1,\ell_2) = e^{-b\ell_2} \delta(\ell_1) 
+ |c| \sqrt{\frac{\ell_2}{\ell_1}}\, e^{-a\ell_1-b\ell_2} I_1\left(2|c|\sqrt{\ell_1\ell_2}\right).
\label{eq:K22-Bessel}
\end{equation}

Equations~\eqref{eq:K12-Bessel}--\eqref{eq:K22-Bessel} show that the
occurrence of modified Bessel functions is not an accidental feature
of particular solvable geometries.  It is the generic outcome of
resumming the switching expansion whenever the inter-target coupling
decomposes into mutually orthogonal two-dimensional invariant
subspaces.

\subsection{Role of geometrical symmetry}

The mode-preserving coupling in Eq.~\eqref{eq:channel-diagonal} arises
naturally in sufficiently symmetric geometries.  For example,
rotational or translational symmetry can decompose the DtN operator
into independent Fourier, angular-momentum, or transverse modes.
Within each mode $\nu$, the two boundary components communicate
without coupling to other modes, leading precisely to a $2\times 2$
matrix from~\eqref{eq:2by2-decomposition}.

For instance, if $\Omega$ is the interval $(0,L)$, its boundary
consists of two endpoints that can naturally be associated with two
targets.  In this special case, the spectrum of the DtN operator has
only two eigenvalues so that $\cM_p$ is actually a $2\times 2$ matrix:
\begin{equation}
\cM_p = \begin{pmatrix} a & c \\ c & b \\ \end{pmatrix}, \qquad a = b = \sqrt{p/D}\, \ctanh(L\sqrt{p/D}), 
\qquad c =  - \frac{\sqrt{p/D}}{\sinh(L\sqrt{p/D})} \,.
\end{equation}
In this case, Eqs.~\eqref{eq:K12-Bessel}--\eqref{eq:K22-Bessel}
provide the exact form of the encounter propagator, as derived in
\cite{Grebenkov2020} by a direct inversion of the double Laplace
transform of $\widetilde G_{q_1,q_2}$.  Indeed,
Eq.~\eqref{eq:P-master} reads as
\begin{equation}  \label{eq:Penc_1d}
\widetilde P(x,\ell_1,\ell_2,p|x_0) = \widetilde G_D(x,p|x_0) \delta(\ell_1) \delta(\ell_2)
+ \frac{1}{D} \sum\limits_{i,i'=1}^2 j_{i,p}(x) j_{i',p}(x_0) K_{ii'}(\ell_1,\ell_2),
\end{equation}
where 
\begin{equation}
j_{1,p}(x) = j_{1,p}(0|x) = \frac{\sinh(x\sqrt{p/D})}{\sinh(L\sqrt{p/D})} \,, \qquad
j_{2,p}(x) = j_{2,p}(L|x) = \frac{\sinh((L-x)\sqrt{p/D})}{\sinh(L\sqrt{p/D})} \,,
\end{equation}
and 
\begin{equation}
\widetilde G_D(x,p|x_0) = \frac{\sinh(x\sqrt{p/D}) \sinh((L-x_0)\sqrt{p/D})}{\sqrt{pD} \sinh(L\sqrt{p/D})} 
\qquad (0 < x \leq x_0 < L).
\end{equation}
We therefore re-derived Eq. (50) from Ref.~\cite{Grebenkov2020}. 

If $\Omega$ is a circular annulus between two concentric circles, both
sets of eigenfunctions $\{\phi_n\}$ and $\{\psi_n\}$ are formed by
Fourier harmonics (see, e.g, \cite{Grebenkov20b}) so that
mode-preserving coupling holds for every $n$.  In this case,
Eqs.~\eqref{eq:K12-Bessel}--\eqref{eq:K22-Bessel} provide the
contribution of each mode to the Laplace-transformed propagator
$\widetilde P(\x,\ell_1,\ell_2,p|\x_0)$, with the coefficients $a$,
$b$ and $c$ being explicitly expressed in terms of modified Bessel
functions.  The same construction holds for a spherical shell between
two concentric spheres (see \cite{Grebenkov20b} for details).

In contrast, for a generic geometry, the off-diagonal block acts
as
\begin{equation*}
C\psi_m = \sum_n c_{nm}\phi_n,
\end{equation*}
so that a switch from one target to the other can simultaneously
change the involved modes.  A typical sequence contributing to the
switching expansion has the form
\begin{equation*}
n_0 \longrightarrow m_1 \longrightarrow n_1 \longrightarrow m_2 \longrightarrow\cdots,
\end{equation*}
with amplitudes containing products such as
\begin{equation*}
c_{n_0m_1} c_{n_1m_1}^* c_{n_1m_2} c_{n_2m_2}^* \cdots.
\end{equation*}
At the same time, the local-time integrations involve several distinct
eigenvalues through the simplex exponential integrals
$\Phi_{m+1}(\ell_1; \alpha_{n_0},\ldots,\alpha_{n_m})$ and their
counterparts on the second target.
There is consequently no single scalar argument whose powers can be
summed into modified Bessel functions $I_0$ or $I_1$, as in
\eqref{eq:K12-Bessel}--\eqref{eq:K22-Bessel}.  The principal
obstruction to the elementary resummation in a generic geometry is
therefore spectral mode mixing induced by the off-diagonal DtN block
$C$.

\subsection{Low-rank inter-target coupling}
\label{sec:low-rank}

Since the off-diagonal DtN block $C$ is smoothing for positively
separated targets, it is compact and its singular values decay to zero
under the above smoothness and separation assumptions.  Their actual
decay rate depends on the boundary regularity and geometry; rapid
decay therefore provides a directly testable criterion for efficiency
of a low-rank approximation:
\begin{equation}
C \simeq C_r = U\Sigma V^\dagger = \sum_{\nu=1}^r \sigma_\nu  |u_\nu\rangle\langle v_\nu|,
\label{eq:lowrank_C}
\end{equation}
where $\{u_\nu\}_{\nu=1}^r\subset L^2(\Gamma_1)$ and
$\{v_\nu\}_{\nu=1}^r\subset L^2(\Gamma_2)$ are orthonormal families,
and $\Sigma=\operatorname{diag}(\sigma_1,\ldots,\sigma_r)$ contains
the retained singular values.  In particular, $C_r^\dagger=V\Sigma
U^\dagger$.

The advantage of this representation becomes transparent in the
renewal formulation.  Introduce the projected semigroups
\begin{equation}
{\sf g}_1(\ell) = U^\dagger e^{-A\ell}U, \qquad
{\sf g}_2(\ell) = V^\dagger e^{-B\ell}V ,
\label{eq:projected_semigroups}
\end{equation}
which are $r\times r$ matrices.  If the spectra
$\{(\alpha_n,\phi_n)\}$ and $\{(\beta_m,\psi_m)\}$ of $A$ and $B$ are
known, these matrices are explicitly given by
\begin{subequations}
\begin{align}
[{\sf g}_1(\ell)]_{\mu\nu} &=  \sum_n e^{-\alpha_n\ell} \langle u_\mu,\phi_n\rangle \langle\phi_n,u_\nu\rangle, \label{eq:g1_spectral}\\
[{\sf g}_2(\ell)]_{\mu\nu} &=  \sum_m e^{-\beta_m\ell} \langle v_\mu,\psi_m\rangle  \langle\psi_m,v_\nu\rangle.  \label{eq:g2_spectral}
\end{align}
\end{subequations}

For instance, replacing $C$ by $C_r$ in the renewal equation for the
first diagonal block gives
\begin{equation}
K_{11} = U_A+T_{11}^{(r)}*K_{11}, \qquad  U_A(\ell_1,\ell_2) =  e^{-A\ell_1}\delta(\ell_2),
\label{eq:lowrank_renewal}
\end{equation}
where $*$ denotes convolution with respect to both local-time
variables, and
\begin{equation}
T_{11}^{(r)}(\ell_1,\ell_2) = e^{-A\ell_1} U\Sigma\,{\sf g}_2(\ell_2)\Sigma U^\dagger .
\label{eq:lowrank_roundtrip}
\end{equation}
Thus,
\begin{equation*}
K_{11} = U_A +T_{11}^{(r)}*U_A +T_{11}^{(r)}*T_{11}^{(r)}*U_A+\cdots .
\end{equation*}

To make the finite-dimensional structure explicit, it is useful to
regard the projected semigroups as kernels of both local-time
variables by setting
\begin{align*}
{\cal E}_1(\ell_1,\ell_2) &=e^{-A\ell_1}U\,\delta(\ell_2),
 & {\cal E}_1^\dagger(\ell_1,\ell_2)
 &=U^\dagger e^{-A\ell_1}\delta(\ell_2),
 \\
{\cal G}_1(\ell_1,\ell_2)
 &={\sf g}_1(\ell_1)\delta(\ell_2),
 & {\cal G}_2(\ell_1,\ell_2)
 &=\delta(\ell_1){\sf g}_2(\ell_2).
\end{align*}
The successive terms of the renewal expansion can then be written
entirely in terms of the two-dimensional convolution as
\begin{align}  \nonumber
K_{11} ={}& U_A
 +{\cal E}_1*\Sigma{\cal G}_2\Sigma*
       {\cal E}_1^\dagger
 +{\cal E}_1*\Sigma{\cal G}_2\Sigma*
       {\cal G}_1*\Sigma{\cal G}_2\Sigma*
       {\cal E}_1^\dagger
 +\cdots  \\
= {}& U_A+
 \sum_{k=1}^{\infty}
 {\cal E}_1*
 \Sigma{\cal G}_2\Sigma*
 \bigl({\cal G}_1*\Sigma{\cal G}_2\Sigma\bigr)^{*(k-1)}
 *{\cal E}_1^\dagger ,
 \label{eq:lowrank_resummed}
\end{align}
where $\bigl({\cal G}_1*\Sigma{\cal
G}_2\Sigma\bigr)^{*0}=I_r\delta(\ell_1)\delta(\ell_2)$.  Thus, the
external factors ${\cal E}_1$ and ${\cal E}_1^\dagger$ retain the full
boundary-space dependence, whereas all intermediate round trips are
described by convolutions of $r\times r$ matrix-valued kernels.

The rank-one case illustrates an important distinction.  If
$C\simeq\sigma|u\rangle\langle v|$, then ${\sf g}_1$ and ${\sf g}_2$
are scalar functions, but $u$ and $v$ need not be eigenfunctions of
$A$ and $B$.  A rank-one coupling therefore does not, by itself, imply
the mode-preserving situation considered above.  Only when the
dominant singular vectors are (approximately) aligned with individual
eigenmodes of $A$ and $B$, does this reduction recover an effective
two-dimensional invariant subspace and the associated Bessel-function
resummation.

\section{Effective reduction for small well-separated targets}
\label{sec:small_targets}

The general operator formulation is simplified considerably when the
two target regions are small compared with the characteristic size of
the confining domain and are separated by a macroscopic distance.  In
this regime, the interaction between the targets is mediated by a
smooth outer field, while the local behavior near each target is
governed by a rescaled inner problem.  This motivates the effective
low-dimensional reduction considered below.  Under an additional
two-mode approximation, the infinite-dimensional block problem reduces
to an effective $2\times2$ problem.  This reduction yields a closed
approximation for the regular part of the encounter propagator and
also clarifies the relative importance of repeated transitions between
the two targets.
We focus here on a three-dimensional domain and consider two targets
$\Gamma_1^\ve$ and $\Gamma_2^\ve$ of characteristic size $\ve \ll 1$,
centered around $\x_1$ and $\x_2$, with surface areas
$S_i=|\Gamma_i^\ve|=O(\ve^2)$.  We assume that
$\operatorname{dist}(\Gamma_1^\ve,\Gamma_2^\ve) =O(1)$.

A related small-target analysis was developed by Bressloff for the
narrow-capture problem with small interior targets \cite{Bressloff22}
(see also \cite{Bressloff21}).  In that work, matched asymptotic
analysis was applied directly to the Laplace-transformed encounter
propagator, but a single boundary local time was assigned to the union
of all targets, thus assuming that all targets obey the same
surface-reaction mechanism.  As noted there, different
target-dependent reaction mechanisms naturally require the
introduction of distinct boundary local times
$\ell_{1,t},\ldots,\ell_{N,t}$.  Our formulation provides such a
multi-local-time generalization and, in the small-target regime, can
be directly compared with the matched-asymptotic results obtained in
Ref.~\cite{Bressloff22} for a single target $\Gamma$.

\subsection{Effective two-mode reduction}
\label{sec:effective-rank-one}

The small-target regime suggests a substantial reduction of the
inter-target coupling, but this reduction should be formulated with
some care.  In particular, for $p>0$ there is in general no reason to
assume that the relevant target mode is approximately constant over
the whole target.  The modified Helmholtz equation introduces the
length scale $L_p=\sqrt{D/p}$, so that the local structure of the
boundary modes may depend on the dimensionless combination
$\ve\sqrt{\frac{p}{D}}$.  For fixed $p$ and $\ve\to0$, this parameter
vanishes and the local problem approaches its $p=0$ counterpart.  This
approximation is, however, not uniform in $p$, and the regime
$p=O(D/\ve^2)$ requires a separate local analysis.

Suppose that the encounter dynamics is dominated by one mode on each
target from the spectral decompositions~\eqref{eq:A-spectrum} of the
two diagonal DtN blocks, namely, $\phi_0^{(p)}$ and $\psi_0^{(p)}$.
Projection of the full block DtN operator onto the two-dimensional
space
$\mathcal H_{\rm eff} = \operatorname{span}
\bigl\{(\phi_0^{(p)},0), (0,\psi_0^{(p)}) \bigr\}$
gives the effective matrix
representation~\eqref{eq:2by2-decomposition}, with
\begin{equation}
a = \alpha_0^{(p)}, \qquad  b = \beta_0^{(p)}, \qquad  c = \langle \phi_0^{(p)}, C\psi_0^{(p)} \rangle = c_{00}(p) .
\label{eq:effective-coefficients}
\end{equation}
The corresponding DtN resolvent restricted to $\mathcal H_{\rm eff}$
is
\begin{equation}
\cR^{\rm eff}_p(q_1,q_2) = 
\begin{pmatrix}
a+q_1 & c\\
c^* & b+q_2
\end{pmatrix}^{-1}
 = \frac{1}{(q_1+a)(q_2+b)-|c|^2}
\begin{pmatrix}
b+q_2 & -c\\
-c^* & a+q_1
\end{pmatrix}.
\label{eq:small-target-resolvent}
\end{equation}

The effective $2\times2$ approximation thus relies on the stronger
property that the dominant inter-target coupling is carried by one
eigenmode of each diagonal DtN block, i.e.,
\begin{equation}
C\psi_0^{(p)} \simeq c\,\phi_0^{(p)}, \qquad C^\dagger\phi_0^{(p)} \simeq c^* \,\psi_0^{(p)},
\label{eq:mode-preserving-rank-one}
\end{equation}
with only subleading projections onto higher modes.
For sufficiently small and well-separated targets, such a reduction is
plausible because the field produced by one target varies smoothly
over the other, whereas the diagonal DtN blocks are controlled by
strongly localized boundary structure.  Nevertheless, the precise
asymptotic validity of Eq.~\eqref{eq:mode-preserving-rank-one}, as
well as its dependence on $p$, requires a separate matched-asymptotic
analysis.
The formulas derived below should therefore be understood as the
closed encounter kernel associated with the effective two-mode
approximation \eqref{eq:2by2-decomposition}.  Their applicability to a
given small-target problem is controlled by the accuracy of the
projection onto $\mathcal H_{\rm eff}$.

\subsection{Closed Bessel approximation}

The double inverse Laplace transform of
Eq.~\eqref{eq:small-target-resolvent} can be performed explicitly.
Introducing $z = 2|c|\sqrt{\ell_1\ell_2}$, one obtains, for
$\ell_1,\ell_2>0$,
\begin{equation}  \label{eq:small-target-K12}
K_{12}^{\rm eff}(\ell_1,\ell_2) =  -c\, e^{-a\ell_1-b\ell_2} I_0(z), \qquad
K_{21}^{\rm eff}(\ell_1,\ell_2) =  -c^*\, e^{-a\ell_1-b\ell_2} I_0(z),
\end{equation}
whereas the regular parts of the diagonal elements are
\begin{equation}  \label{eq:small-target-K11}
K_{11}^{\rm eff,reg}(\ell_1,\ell_2) = |c| \sqrt{\frac{\ell_1}{\ell_2}}\, e^{-a\ell_1-b\ell_2} I_1(z), \qquad
K_{22}^{\rm eff,reg}(\ell_1,\ell_2) = |c| \sqrt{\frac{\ell_2}{\ell_1}}\, e^{-a\ell_1-b\ell_2} I_1(z).
\end{equation}
The singular parts, $e^{-a\ell_1}\delta(\ell_2)$ and
$\delta(\ell_1)e^{-b\ell_2}$, correspond to the cases when only one of
the two targets has been encountered and are therefore not included in
Eqs.~\eqref{eq:small-target-K11}. 

Substitution into the general representation~\eqref{eq:P-master} of
the encounter propagator gives the following closed approximation for
its regular part (with $\ell_1,~\ell_2>0$):
\begin{align}
\widetilde P_{11}(\x,\ell_1,\ell_2,p|\x_0) \simeq &
\frac{e^{-a\ell_1-b\ell_2}}{D}
\Bigg\{|c|I_1(z)\bigg[\sqrt{\frac{\ell_1}{\ell_2}}\, J_1^{\rm out}(p|\x) J_1^{\rm in}(p|\x_0)
+ \sqrt{\frac{\ell_2}{\ell_1}}\, J_2^{\rm out}(p|\x) J_2^{\rm in}(p|\x_0)\bigg]
\nonumber\\
&\qquad \qquad
-I_0(z)\bigg[c\, J_1^{\rm out}(p|\x) J_2^{\rm in}(p|\x_0)
+ c^*\, J_2^{\rm out}(p|\x) J_1^{\rm in}(p|\x_0)\bigg] \Bigg\},
\label{eq:small-target-P11-Bessel}
\end{align}
where we introduced the projected boundary flux amplitudes:
\begin{subequations}
\begin{align}
J_1^{\rm in}(p|\x_0) &= \left\langle \phi_0, j_{1,p}(\cdot|\x_0) \right\rangle , \qquad
J_2^{\rm in}(p|\x_0) = \left\langle \psi_0, j_{2,p}(\cdot|\x_0) \right\rangle , \\
J_1^{\rm out}(p|\x) &= \left\langle j_{1,p}(\cdot|\x), \phi_0 \right\rangle , \qquad
J_2^{\rm out}(p|\x) = \left\langle j_{2,p}(\cdot|\x), \psi_0 \right\rangle .
\label{eq:projected-fluxes}
\end{align}
\end{subequations}
Equation~\eqref{eq:small-target-P11-Bessel} is the closed
approximation resulting from the effective two-mode reduction.  It has
the same modified-Bessel-function structure as the exact expressions
like~\eqref{eq:Penc_1d}, obtained in elementary geometries but its
coefficients are now determined by the local target properties and by
propagation through an otherwise arbitrary outer domain.

It is instructive to rewrite the
approximation~\eqref{eq:small-target-P11-Bessel} in a more transparent
form.  First, we recall that the parameter $c$ is real whenever $p
\geq 0$ (see the related discussion in Sec. \ref{sec:DtN_def}).
Second, we note that even though the signs of the normalized
eigenfunctions $\phi_0$ and $\psi_0$ are arbitrary,
Eq.~\eqref{eq:small-target-P11-Bessel} is invariant under their
independent sign changes because the coupling $c$ and the projected
flux amplitudes transform simultaneously.  We henceforth fix this
ambiguity by choosing the principal eigenfunctions $\phi_0$ and
$\psi_0$ positive on their respective targets.  With this convention,
the maximum principle and Hopf's lemma imply $C \psi_0<0$ on
$\Gamma_1$ under the usual connectedness and regularity assumptions,
and therefore $c=\langle\phi_0, C\psi_0\rangle<0$.  Defining $\gamma =
-c$, Eq.~\eqref{eq:small-target-P11-Bessel} takes the manifestly
additive form
\begin{align}
D \widetilde P_{11}(\x,\ell_1,\ell_2,p|\x_0) \simeq \gamma \, 
e^{-a\ell_1-b\ell_2} & 
\Bigg\{ I_0\left(2\gamma \sqrt{\ell_1\ell_2}\right)
\big[J_1^{\rm out}(p|\x)J_2^{\rm in}(p|\x) + J_2^{\rm out}(p|\x)J_1^{\rm in}(p|\x)\big]
\nonumber\\
&+
I_1\left(2\gamma\sqrt{\ell_1\ell_2}\right)
\bigg[\sqrt{\frac{\ell_1}{\ell_2}}\, J_1^{\rm out}(p|\x) J_1^{\rm in}(p|\x)
+ \sqrt{\frac{\ell_2}{\ell_1}}\, J_2^{\rm out}(p|\x) J_2^{\rm in}(p|\x) \bigg]\Bigg\} .
\label{eq:small-target-P11-gamma}
\end{align}

\subsection{Relation to the common-local-time asymptotics}
\label{sec:common-local-time}

There is a direct relation between the present two-local-time
representation and the matched-asymptotic formulas obtained when one
assigns a single boundary local time to the union of the targets:
$\ell_t = \ell_{1,t} + \ell_{2,t}$; as a consequence, the encounter
propagator associated with the total boundary local time $\ell_t$ is
that
\begin{equation}
P_{\rm tot}(\x,\ell,t|\x_0) = \int\limits_0^\ell d\ell_1 \, P(\x,\ell_1,\ell-\ell_1,t|\x_0).
\end{equation}
In Ref.~\cite{Bressloff22}, the first inter-target correction to the
encounter propagator $\widetilde P_{\rm tot}(\x,\ell,p|\x_0)$ for two
spherical targets of radii $\ve \rho_1$ and $\ve \rho_2$ contains
expressions of the form
\begin{equation}
\frac{e^{-\widehat\ell/\rho_2} - e^{-\widehat\ell/\rho_1}}{\rho_1^{-1}-\rho_2^{-1}} \,,
\label{eq:Bressloff-difference}
\end{equation}
where $\widehat\ell=\ell/\varepsilon$ is the rescaled total local
time.  This structure has a simple interpretation within the present
formalism.
Indeed, the lowest one-switch contribution for two distinct local
times has the form $K_{12}^{(1)}(\ell_1,\ell_2) \propto
e^{-a\ell_1-b\ell_2}$.  If only the total local time
$\ell=\ell_1+\ell_2$ is retained, the separate contributions of
$\ell_1$ and $\ell_2$ have to be integrated along the line
$\ell_1+\ell_2=\ell$, yielding
\begin{align}
\int_0^\ell d\ell_1\, e^{-a\ell_1} e^{-b(\ell-\ell_1)} &= \frac{e^{-b\ell}-e^{-a\ell}}{a-b} \,,  \qquad a\neq b
\label{eq:marginal-convolution}
\end{align}
(and $\ell e^{-a\ell}$ for $a = b$).  For $a=(\ve\rho_1)^{-1}$ and
$b=(\ve\rho_2)^{-1}$ as the leading-order terms of the principal
eigenvalues $\alpha_0^{(p)}$ and $\beta_0^{(p)}$ for spherical
targets, Eq.~\eqref{eq:marginal-convolution} reduces, after the
rescaling $\widehat\ell=\ell/\ve$, to
Eq.~\eqref{eq:Bressloff-difference}.
Thus the simplex exponential integrals appearing in the
total-local-time matched-asymptotic expansion can be interpreted as
marginals of the simpler product structure of the two-local-time
propagator.  Conversely, resolving the total boundary local time into
target-specific components removes the convolution and replaces the
difference quotient by the product $e^{-a\ell_1-b\ell_2}$.
This correspondence provides a useful consistency check of the
switching expansion and clarifies the relation between the present
multi-local-time DtN formulation and the narrow-target
matched-asymptotic approach from~\cite{Bressloff22}.

\subsection{Scaling of the effective DtN coefficients}
\label{sec:scaling-effective-dtn}

The scaling of the coefficients $a$, $b$ and $c$ entering the
effective matrix $\mathcal M_p^{\rm eff}$
in~\eqref{eq:2by2-decomposition} requires some care because the
small-$\ve$ asymptotic behavior may not hold uniformly up to $p = 0$.
Let us clarify this issue by first considering the small-target limit
at fixed $p>0$.

Let $\Gamma_i^\ve = \x_i + \ve\widehat\Gamma_i$, where
$\widehat\Gamma_i$ is a fixed rescaled target.  In the inner variables
$\x = \x_i + \ve\y$, the modified Helmholtz equation becomes
$\left(-\Delta_{\bm y} + p\ve^2/D \right)u=0$.  Hence, whenever
$\ve\sqrt{p/D} \ll 1$, the leading inner problem is harmonic and
independent of $p$.  This scaling suggests
\begin{equation}
a = \frac{\widehat a}{\ve}+o(\ve^{-1}), \qquad  b = \frac{\widehat b}{\ve} + o(\ve^{-1}),
\label{eq:small-target-ai-scaling}
\end{equation}
where $\widehat a>0$ and $\widehat b>0$ depend only on the local shape
of the corresponding rescaled target.  
The scaling~\eqref{eq:small-target-ai-scaling} is consistent with the
matched-asymptotic encounter formulation of Ref.~\cite{Bressloff22}.
For a spherical interior target of radius $r_j=\ve\rho_j$, the
local-time variable is rescaled as $\widehat\ell=\ell/\ve$, and the
leading inner solution contains the factor $\exp(-\widehat\ell/\rho_j)
= \exp(-\ell/r_j)$.  Thus, for a spherical target, the leading
local-time decay rate is $r_j^{-1}=O(\ve^{-1})$, while the
characteristic accumulated local time is $O(\ve)$.  This provides a
direct matched-asymptotic counterpart of the scaling $a,b=O(\ve^{-1})$
reported above.
The numerical coefficient is, however, geometry dependent.  For an
interior sphere it is exactly $1/r_j$ at leading order, whereas for a
small target located on the boundary of a confining domain the
corresponding coefficient is determined by a different local problem
in the tangent half-space.  We therefore retain the more general
notations $\widehat a/\ve$ and $\widehat b/\ve$.
A matched-asymptotic analysis of the DtN eigenvalues and
eigenfunctions for a single target and $p>0$ was developed in
Ref. \cite{Grebenkov25}, while Refs. \cite{Grebenkov26a,Grebenkov26b}
address multi-target settings at $p=0$.  The leading $O(\ve^{-1})$
scaling in Eq.~\eqref{eq:small-target-ai-scaling}, however, already
follows from the elementary inner rescaling above.

The off-diagonal block has a different origin.  Since the two targets
are separated by an $O(1)$ distance, the field generated by one target
is smooth on the scale of the other.  Its leading contribution is
therefore expected to be determined by the monopolar parts of the two
local problems.  This suggests the leading structure
\begin{equation}
c \simeq  -\mathcal Q_1\mathcal Q_2 \, D\widetilde G_N(\x_1,p|\x_2),
\label{eq:c-green}
\end{equation}
where $\mathcal Q_i$ are coupling amplitudes determined by the local
target problems, while $\widetilde G_N(\x,p|\x_0)$ is the Neumann
Green's function of the outer modified Helmholtz equation satisfying
\begin{equation}
(p-D\Delta)\widetilde G_N(\x,p|\x_0) = \delta(\x-\x_0), \quad \x\in\Omega,
\qquad  \partial_n\widetilde G_N=0
\quad\mbox{on }\partial\Omega .
\end{equation}
A matched-asymptotic calculation would be needed to determine the
coefficients $\mathcal Q_i$ and higher-order corrections in
Eq.~\eqref{eq:c-green}.
This structure of the off-diagonal coupling also appears explicitly in
the matched-asymptotic analysis of the total boundary local time
in~\cite{Bressloff22} for the case of well-separated spherical
targets.  For fixed $p>0$ and an $O(1)$ separation between the
targets, $D\widetilde G_N(\x_1,p|\x_2)=O(1)$, which, together with the
monopolar coupling assumption~\eqref{eq:c-green}, implies $c = O(1)$.
In summary, we have for fixed $p>0$
\begin{equation}
a=O(\ve^{-1}), \qquad b=O(\ve^{-1}), \qquad c=O(1).
\label{eq:fixed-p-hierarchy}
\end{equation}
Thus repeated inter-target transfers are parametrically weaker than
the local DtN dynamics on each target.

Importantly, Eq.~\eqref{eq:fixed-p-hierarchy} is not uniform as
$p\to0$.  The Neumann Green's function contains the spatially uniform
mode,
\begin{equation}
D \widetilde G_N(\x,p|\x_0) = \frac{D}{p|\Omega|} + G_N(\x,\x_0) + O(p),
\label{eq:small-p-neumann-green}
\end{equation}
where $G_N$ is the reduced Neumann Green's function
\cite{Grebenkov25}.  Consequently, the fixed-$p$ estimate of the
inter-target coupling ceases to be parametrically smaller when $p$
approaches the narrow-target scale.  Schematically, the crossover
occurs at $p=O\left(D\ve/|\Omega|\right)$.
The breakdown of the fixed-$p$ hierarchy as $p$ approaches the
narrow-target scale is also consistent with the exact $p=0$ structure.
Indeed, the DtN operator $\cM_0$ has a zero eigenvalue associated with
a constant eigenfunction on $\Gamma_1\cup\Gamma_2$.  The effective
off-diagonal coupling therefore cannot remain $O(1)$ relative to
$a,b=O(\ve^{-1})$ all the way to $p=0$.
The fixed-$p$ hierarchy \eqref{eq:fixed-p-hierarchy} should thus be
understood as an intermediate asymptotic regime.  This suggests the
schematic scaling window $\frac{D\ve}{|\Omega|} \ll p \ll
\frac{D}{\ve^2}$, where geometry-dependent dimensionless factors have
been omitted.  For every fixed $p>0$, this window is eventually
reached as $\ve\to0$.

\subsection{Fixed-$p$ narrow-target limit}

The scaling $\ell_i=O(\ve)$ follows both from
Eq.~\eqref{eq:small-target-ai-scaling} and from the local
matched-asymptotic encounter problem.  For a spherical target,
Ref.~\cite{Bressloff22} obtains the leading dependence
$\exp[-\ell_i/(\ve\rho_i)]$, which shows directly that the natural
rescaled local time is $\widehat\ell_i = \ell_i/\ve =O(1)$.  We adopt
the same scaling for the generic effective two-mode problem.  On this
scale, $a\ell_1=O(1)$ and $b\ell_2=O(1)$, so that the exponential
factors in Eq.~\eqref{eq:small-target-P11-Bessel} remain nontrivial.

On the other hand, Eq.~\eqref{eq:fixed-p-hierarchy} implies $z =
2\gamma\sqrt{\ell_1\ell_2} = O(\ve)$, with $\gamma = -c > 0$.  Thus
the modified Bessel functions themselves admit a regular small-$\ve$
expansion: $I_0(z) = 1 + O(\ve^2)$ and $I_1(z) = z + O(\ve^3)$.  This
observation leads to an additional simplification of the rank-one
Bessel approximation: the off-diagonal kernels become
\begin{equation*} 
K_{12}^{\rm eff} = \gamma \, e^{-a\ell_1-b\ell_2} \left[1+O(\ve^2) \right], \qquad
K_{21}^{\rm eff} = \gamma \, e^{-a\ell_1-b\ell_2} \left[1+O(\ve^2)\right],
\end{equation*}
whereas the diagonal regular contributions are
\begin{equation*} 
K_{11}^{\rm eff,reg} = \gamma^2\ell_1 e^{-a\ell_1-b\ell_2} \left[1+O(\ve^2)\right], \qquad
K_{22}^{\rm eff,reg} = \gamma^2\ell_2 e^{-a\ell_1-b\ell_2} \left[1+O(\ve^2)\right].
\end{equation*}
Consequently, the dominant regular contribution to the encounter
propagator is
\begin{equation}
D \widetilde P_{11} \simeq  \gamma\, e^{-a\ell_1-b\ell_2}
\bigg[J_1^{\rm out}(p|\bm{x}) J_2^{\rm in}(p|\bm{x}_0)
+ J_2^{\rm out}(p|\bm{x}) J_1^{\rm in}(p|\bm{x}_0) \bigg].
\label{eq:small-target-P11-leading}
\end{equation}
This expression has a simple interpretation.  A trajectory
contributing to $\widetilde P_{11}$ with $\ell_1>0$, $\ell_2>0$ must
encounter both targets.  In the fixed-$p$ narrow-target limit, within
the effective two-mode approximation, the dominant contribution
corresponds to a single inter-target transition, i.e., to the two
possible sequences: $\Gamma_1\longrightarrow\Gamma_2$ and
$\Gamma_2\longrightarrow\Gamma_1$, which produce the two terms in
Eq.~\eqref{eq:small-target-P11-leading}.
The first correction to Eq.~\eqref{eq:small-target-P11-leading} arises
from the diagonal regular kernels and corresponds to two inter-target
transfers:
\begin{equation}
D \widetilde P_{11} \simeq  \gamma \, e^{-a\ell_1-b\ell_2}
\Big[J_1^{\rm out}J_2^{\rm in} + J_2^{\rm out}J_1^{\rm in}
+ \left(\gamma \ell_1 J_1^{\rm out}J_1^{\rm in} + \gamma \ell_2 J_2^{\rm out}J_2^{\rm in}\right) + \ldots \Big].
\label{eq:small-target-P11-next}
\end{equation}
Because $\ell_i=O(\ve)$, the two-switch terms are smaller by one power
of $\ve$ than the one-switch contribution, while additional round
trips are progressively suppressed.

\section{Discussion and conclusion}
\label{sec:discussion}

We have developed a spectral formulation of the encounter propagator
for restricted diffusion in a bounded domain with multiple targets, in
which boundary local times are resolved separately.  The main step
consisted in performing a multidimensional Laplace transform inversion
of the multi-parameter DtN resolvent $(\mathcal M_p+q_1P_1+\cdots +
q_NP_N)^{-1}$ with respect to the parameters $q_1,\ldots,q_N$,
quantifying reactivities of different targets.  The essential
difficulty as compared to the single-target problem is the
noncommutativity of the DtN operator with the target projection
operators $P_i$, so that the multi-target encounter problem is
intrinsically matrix-valued.  The block-DtN representation resolved
this difficulty: the local-time inversions become ordinary semigroup
evolutions generated by the diagonal DtN blocks, while the
off-diagonal blocks produce transfers between targets.
This representation yields an absolutely convergent switching
expansion, which possesses an encounter interpretation in terms of
successive switches between the target regions.  We also discussed an
equivalent operator-valued renewal equation.  In target-restricted DtN
eigenbases, the encounter kernel was determined by ordinary DtN-block
spectra and an inter-target coupling matrix.  In particular, the
natural spectral data for the two-target encounter problem can be
summarized as $\left\{ \alpha_n^{(p)},\phi_n^{(p)}; \quad
\beta_m^{(p)},\psi_m^{(p)}; \quad c_{nm}(p) \right\}$.  When this
coupling preserves spectral modes, the series resums exactly into
modified Bessel functions, thereby explaining the structure previously
observed in elementary geometries.  For generic separated targets, the
smoothing character of the off-diagonal coupling instead suggested
rapid spectral convergence and low-rank matrix approximations.

The singular decomposition in Eq.~\eqref{eq:P-decomposition} is also
structurally important.  It separates paths that have encountered
neither target, exactly one target, or both targets.  This
decomposition should be useful for deriving marginal local-time
distributions, moments and correlations of $\ell_{1,t}$ and
$\ell_{2,t}$, and reaction observables associated with arbitrary
stopping conditions imposed on the two boundary local times.  While we
focused on the two-target setting, the construction extends directly
to any finite number of target regions.  It thus provides a
generic-domain framework in which geometry and target-specific
encounter statistics are separated at the operator level, opening a
route to describe joint local-time distributions, target competition,
and diffusion-controlled reactions with distinct surface mechanisms.

For small and well-separated targets, we further examined an effective
two-mode reduction.  At fixed $p>0$, the local DtN scales and the
smooth inter-target coupling lead to a regime in which repeated
transfers are progressively suppressed, with the leading joint
contribution associated with a single transfer between the targets.  A
systematic matched-asymptotic analysis could further quantify the
accuracy of this reduction, while numerical tests of the
singular-value decay of the inter-target coupling operator $C$ would
assess the efficiency of low-rank approximations in generic
geometries.

The present construction thus identifies a hierarchy of descriptions
according to the structure of the inter-target DtN coupling: (i)
Generic separated targets lead to an operator-valued renewal problem;
(ii) rapidly decaying singular values permit low-rank matrix
approximations; (iii) mode-preserving coupling yields independent
two-dimensional problems and exact Bessel resummation; and, (iv)
within an effective two-mode description of small targets at fixed
$p$, the leading contribution reduces further to a single inter-target
transfer.  This hierarchy connects generic geometries to the explicit
formulas known in separable domains.

Several questions deserve further analysis, including the asymptotic
decay of the spectral coupling coefficients $c_{nm}(p)$, the
convergence rate of finite-dimensional truncations, the limit of
nearby or touching targets, and efficient numerical inversion with
respect to $p$.  The block structure may also provide systematic
asymptotic expansions for well-separated small targets.  It would be
particularly interesting to combine the block-DtN formulation
developed here with the matched-asymptotic construction of
Ref.~\cite{Bressloff22}.  The latter treats the total boundary local
time for several small targets, whereas the present theory resolves
the local times accumulated on each target separately.  The
convolution identity discussed in Sec.~\ref{sec:common-local-time}
suggests that the two descriptions are complementary and can be
matched order by order in the target size.

\section*{Acknowledgment}

The author used ChatGPT (OpenAI) for language editing and assistance
in exploring some analytical derivations.  The author assumes
responsibility for all content.

\end{document}